\documentclass[10pt,journal]{IEEEtran}

\renewcommand{\IEEEQED}{\IEEEQEDclosed}
\usepackage{color}

\usepackage{}
\usepackage{subfigure}
\usepackage{graphicx,cite,epsfig,amssymb,amsmath,multirow,lettrine,flushend,extarrows}
\usepackage{algorithm}
\usepackage{algorithmic}

\begin{document}
\bibliographystyle{ieeetr}

%------------------------------- title ----------------------------------------

\title{Fog-Assisted Operational Cost Reduction \\ for Cloud Data Centers}

\author{{Liang~Yu,~\IEEEmembership{Member,~IEEE}, Tao~Jiang,~\IEEEmembership{Senior Member,~IEEE}, and Yulong Zou,~\IEEEmembership{Senior Member,~IEEE}}
\thanks{
\newline L. Yu and Y. Zou are with Key Laboratory of Broadband Wireless Communication and Sensor Network Technology of Ministry of Education, Nanjing University of Posts and Telecommunications, Nanjing 210003, P. R. China. \newline
T. Jiang is Wuhan National Laboratory for Optoelectronics, School of Electronics Information and Communications, Huazhong University of Science and Technology, Wuhan 430074, P. R. China. \newline
}}

%\markboth{IEEE Access, Vol. XX, No. XX, 2017} {{et al.}: Fog-Assisted Operational Cost Reduction for Cloud Data Centers}
\maketitle

\maketitle

\begin{abstract}
In this paper, we intend to reduce the operational cost of cloud data centers with the help of fog devices, which can avoid the revenue loss due to wide-area network propagation delay and save network bandwidth cost by serving nearby cloud users. Since fog devices may not be owned by a cloud service provider, they should be compensated for serving the requests of cloud users. When taking economical compensation into consideration, the optimal number of requests processed locally by each fog device should be decided. As a result, existing load balancing schemes developed for cloud data centers can not be applied directly and it is very necessary to redesign a cost-ware load balancing algorithm for the fog-cloud system. To achieve the above aim, we first formulate a fog-assisted operational cost minimization problem for the cloud service provider. Then, we design a parallel and distributed load balancing algorithm with low computational complexity based on Proximal Jacobian Alternating Direction Method of Multipliers (PJ-ADMM). Finally, extensive simulation results show the effectiveness of the proposed algorithm.
\end{abstract}

\section{Introduction}\label{s1}

Cloud computing is envisioned as an effective means of providing worldwide consumers with on-demand computing resources (e.g., networks, servers, storage, applications, and services) in a convenient way. Due to the advantages of high resource utilization, strong computing ability, high reliability, and rapid elasticity of cloud computing, many Internet workloads are processed in cloud data centers. According to a report in Cisco Global Cloud Index (2015-2020), 92 percent of all data center workloads are expected to be processed in cloud data centers by 2020\cite{Ciso2015}. Since the operational cost of a cloud data center is very high (e.g., as a significant fraction of the operational cost, the annual energy cost related to many cloud service providers was larger than millions of dollars\cite{Qureshi2009}), it is of great importance to reduce the operational cost for a cloud service provider.

There have been lots of schemes on reducing the operational cost/energy cost of cloud data centers \cite{Greenberg2009,Ghamkhari2016,Gao2012,Lei2012,Guo2013,Xu2014,liangyu2015,LiangArxiv2016,Islam2015,Tran2016}, such as dynamic server provisioning, spatial/temporal load balancing, energy storage, incorporating renewable energies, partial execution, and participating in demand response programs of smart grids. In \cite{Lei2012}, Rao \emph{et al.} proposed a geographical load balancing (GLB) scheme to minimize the energy cost of data centers in deregulated electricity markets. In \cite{Islam2015}, Ren \emph{et al.} presented a water-constrained GLB scheme to minimize the operational cost of data centers. In \cite{Guo2013}, Guo \emph{et al.} proposed an energy cost saving strategy for data centers using energy storage. In \cite{Xu2014}, Xu \emph{et al.} investigated the problem of reducing the peak power demand and energy cost of data centers using partial execution. In \cite{liangyu2015}\cite{LiangArxiv2016}, Yu \emph{et al.} studied the problem of reducing the energy/operational cost for geo-distributed data centers in smart microgrids with the consideration of electricity selling/buying, energy storage, load balancing, renewable energies, and dynamic server provisioning or partial execution. In addition, the operational cost of cloud data centers could be offset partially by economical compensation obtained from the participation of demand response programs\cite{Tran2016}.

Different from existing schemes, we intend to reduce the operational cost of cloud data centers with the help of fog devices\cite{Mchiang2016,Bonomi2012}, which are capable of offering certain advantages by serving nearby cloud users, e.g., avoiding wide-area network (WAN) propagation delay and reducing network bandwidth cost. Since fog devices may not be owned by a cloud service provider\cite{Mchiang2016}, the cloud service provider should compensate them for their efforts in serving user requests. When taking economical compensation into consideration, the optimal number of requests processed locally by each fog device should be decided. As a result, existing load balancing schemes developed for cloud data centers can not be applied directly and it is necessary to redesign a cost-ware load balancing algorithm for the fog-cloud system. To achieve the above aim, we first formulate a fog-assisted operational cost minimization problem for the cloud service provider, which is a large-scale mixed integer linear programming (MILP). To solve the formulated problem efficiently, we propose a parallel and distributed load balancing algorithm with low computational complexity based on Proximal Jacobian Alternating Direction Method of Multipliers (PJ-ADMM)\cite{Deng2014}. Though ADMM-based algorithm has been developed in \cite{Xu2014} for the operational cost reduction of cloud data centers, it could not be directly used in our problem since standard ADMM is applicable to the convex optimization problem with two-block variables, while there are four-block variables in our optimization problem.

The contributions of this paper could be summarized below:
\begin{itemize}
  \item We formulate a fog-assisted operational cost minimization problem for a cloud service provider, where the cost consists of four parts, namely the energy cost of cloud data centers, network bandwidth cost, revenue loss due to the WAN propagation delay, and the economic compensation paid to fog devices.
  \item We propose a parallel and distributed algorithm for the formulated problem based on PJ-ADMM. Note that the proposed algorithm has low computational complexity since all decisions could be made based on close-form expressions or binary search.
  \item Extensive simulations show that the proposed algorithm could help the cloud to save operational cost effectively in the presence of fog devices.
\end{itemize}

The rest of this paper is organized as follows. Section \ref{s2} introduces the system model and the problem formulation. Section \ref{s3} proposes a distributed algorithm based on PJ-ADMM. Then, we conduct extensive simulations in Section \ref{s4}. Finally, conclusions are made in Section \ref{s5}.

\section{Model And Formulation}\label{s2}
\begin{figure}[!htb]
\centering
\includegraphics[scale=0.4]{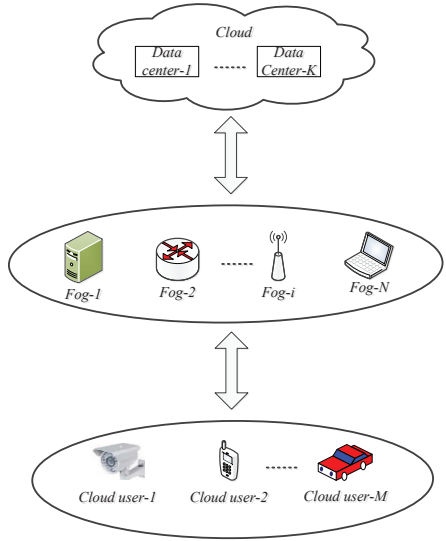}
\caption{System model}\label{fig_1}
\end{figure}

The system model studied in this paper is shown in Fig.~\ref{fig_1}, where the cloud provides service to its cloud users. To improve the user experience and reduce network bandwidth cost, fog devices (e.g., routers, servers, laptops) in the vicinity of cloud users could be selected as helpers of the cloud. In return, fog devices would receive economical compensation for their efforts. Suppose there are $N$ available fog devices and $K$ geographically distributed cloud data centers in the considered fog-cloud system. To serve $J$ types of requests (this paper mainly focuses on delay-sensitive requests) from nearby cloud users, each fog device $i$ ($1\leq i\leq N$) should decide the type and quantity of application requests to be processed. Then, the remaining requests are dispatched to cloud data centers. In the following parts, we would provide the models related to workload allocation, power consumption, operational cost. Then, we formulate a fog-assisted operational cost minimization problem, which is solved periodically at the beginning of each time slot, e.g., every 1 hour.

\subsection{Workload Allocation Model}
When the fog device $i$ receives the application requests from cloud users, it will process part of them locally and dispatch the remaining requests to the remote cloud. Let $\alpha_{i,j}$ be the request rate of application $j$ allocated to the fog device $i$ (in requests/second) and $v_{i,j}$ be the service rate of application $j$ supported by fog device $i$ (in Mbps), we have\cite{DengR2016}
\begin{align} \label{f_1}
&\frac{1}{v_{i,j}/s_j-\alpha_{i,j}}\leq t_j^{\max},~\forall~i,j,\\
&~~~~~\alpha_{i,j}\geq 0,~\forall~i,k,
\end{align}
where $s_j$ is the request size of application $j$ (in Mb/request) and $t_j^{\max}$ (in seconds) is the maximum tolerant delay of application $j$.

Let $\lambda_{i,j}$ ($1\leq j\leq J$) and $\beta_{i,j,k}$ ($1\leq k\leq K$) denote the request rate of application $j$ arriving at fog device $i$ and the request rate of application $j$ allocated from fog device $i$ to data center $k$ (both in requests/second), respectively. According to workload balance, we obtain
\begin{align} \label{f_2}
\alpha_{i,j}+\sum\nolimits_{k} \beta_{i,j,k}=\lambda_{i,j},~\forall~i,j.
\end{align}

Since the ISP link capacities of cloud data centers are limited, we have the following constraints,
\begin{align} \label{f_3}
&\sum\nolimits_i\sum\nolimits_{j} \beta_{i,j,k}s_j\leq A_k^{\max},~\forall~k,\\
&~~~~~\beta_{i,j,k}\geq 0,~\forall~i,j.
\end{align}
where $A_k^{\max}$ is the link capacity of data center $k$ (in Mbps).

In cloud data centers, the number of active servers could be dynamically configured to ensure that the requests of application $j$ could be finished within the maximum tolerant delay $t_j^{\max}$. Then, we have\cite{liangyu2015}
\begin{align} \label{f_4}
&\frac{1}{c_{j,k}\mu_{j,k}-\sum\nolimits_i\beta_{i,j,k}}+\frac{1}{\mu_{j,k}}\leq t_j^{\max},~\forall~k,\\
&~~~~~~~~~~0\leq c_{j,k}\leq C_{j,k},~\forall~j,m, \\
&~~~~~~~~~~~~~~~c_{j,k}\in \mathbb{N}^+,
\end{align}
where $\mu_{j,k}$ denotes the service rate of servers for application $j$ in data center $k$ (in requests/second); $c_{j,k}$ and $C_{j,k}$ are the number of active servers and the total number of servers for application $j$ in data center $k$, respectively.

\subsection{Power Consumption Models Associated with Cloud Data Centers and Fog Devices}
Let $\text{PUE}_k$ denote the power usage effectiveness of the data center $k$. In addition, denote the idle power and peak power of the servers for application $j$ in data center $k$ by $p_{j,k}^{\text{idle}}$ and $p_{j,k}^{\text{peak}}$ (both in Watts), respectively. Then, the power consumption of data center $k$ could be estimated by\cite{Qureshi2009}
\begin{align} \label{f_5}
P_k^{\text{cloud}}=\sum\nolimits_{j}\big(c_{j,k}a_{j,k}+b_{j,k}\frac{\sum\nolimits_i \beta_{i,j,k}}{\mu_{j,k}}\big),
\end{align}
where $a_{j,k}=p_{j,k}^{\text{idle}} + (\text{PUE}_k - 1)p_{j,k}^{\text{peak}}$; $b_{j,k}=p_{j,k}^{\text{peak}} - p_{j,k}^{\text{idle}}$.

For fog devices, their power consumptions ($P_i^{\text{fog}}$) could be calculated according to a linear energy consumption model as in \cite{Jalali2016}\footnote{Note that different energy consumption models of fog devices would not affect the nature of the investigated problem. Moreover, the proposed algorithm is applicable on the condition that the energy consumption models are convex, e.g., quadratic power consumption function in \cite{DengR2016}.},
\begin{align} \label{f_6}
P_i^{\text{fog}}=\Big(q_{i}^\text{idle}+(q_{i}^\text{peak}-q_{i}^\text{idle})\frac{\sum\nolimits_j \alpha_{i,j}s_j}{v_{i}}\Big),~\forall~i,
\end{align}
where $v_i=\sum\nolimits_j v_{i,j}$; $q_{i}^\text{idle}$ and $q_{i}^\text{peak}$ are the idle power and peak power of fog device $i$, respectively.

\subsection{Operational Cost Model}
In this work, we focus on minimizing the operational cost of a cloud service provider in a given time slot with duration $T$ without sacrificing the interests of fog devices, where the cost consists of four parts, i.e., the energy cost of data centers, network bandwidth cost, revenue loss associated with WAN propagation delay, and the economic compensation paid to the fog devices.

Let $\nu_k$ denote the electricity price associated with electric region that data center $k$ is located. Then, the energy cost of all data centers is given by
\begin{align} \label{f_7}
\Gamma_1=\sum\nolimits_j \sum\nolimits_k\nu_k\big(c_{j,k}a_{j,k}+b_{j,k}\frac{\sum\nolimits_i \beta_{i,j,k}}{\mu_{j,k}}\big)T.
\end{align}

When receiving a user request, cloud data centers should process it and give a response packet to the user. In reality, both request transmitting and response packet could generate the traffic. Since the traffic volume of user requests is usually far smaller than that of responses, we mainly focus on the traffic generated by response and assume that the traffic volume generated by response for a request of application $j$ is $\tau_j$. Let $B_k$ be the bandwidth price of the ISP link connected to data center $k$ (in \$/Mbps/month). Then, the network bandwidth cost is given by\cite{Xu2013}
 \begin{align} \label{f_8}
\Gamma_2=\sum\nolimits_i\sum\nolimits_j\sum\nolimits_k\beta_{i,j,k}\tau_jB_k.
\end{align}

For delay-sensitive requests, a moderate increase in user-perceived latency would result in substantial revenue loss for the cloud service provider\cite{Greenberg2009}. Compared with the case that all incoming requests are processed at fog devices, transmitting requests from fog devices to data centers would result in extra WAN propagation latency. Denote the propagation latency of the requests associated with fog device $i$ and data center $k$ by $L_{i,k}$ (in ms), which could be measured through empirical approaches\cite{Xu2013}. Then, the revenue loss due to the WAN propagation latency is obtained by\cite{LiangArxiv2016},
\begin{align} \label{f_9}
\Gamma_3=\sum\nolimits_i\sum\nolimits_j\sum\nolimits_k\omega_j L_{i,k}\beta_{i,j,k}T,
\end{align}
where $\omega_j$ denotes the latency conversion parameter that translates network propagation latency into revenue loss of application $j$ (in \$/ms/req).

Since fog devices could help the cloud to reduce WAN propagation latency and bandwidth cost, the cloud service provider should make a compensation to fog devices for their energy costs incurred by serving requests. We assume that compensations obtained by fog devices are proportional to the served requests. The reason behind this assumption is that fog devices are willing to serve the requests for the cloud service provider on the condition that their interests would not be damaged. According to the above assumption, the compensation paid to the fog devices is given by
\begin{align} \label{f_10}
\Gamma_4=\sum\nolimits_iS_ih_iq_{i}\frac{\sum\nolimits_j\alpha_{i,j}s_j}{v_{i}}T,
\end{align}
where $h_i\geq 1$ is the compensation factor associated with fog device $i$ and $S_i$ is the electricity price associated with electric region that fog device $i$ is located; $q_{i}=q_{i}^\text{peak}-q_{i}^\text{idle}$. Note that we implicitly neglects the economical compensation for the fixed energy consumption of fog devices since serving cloud users are not necessarily their sole purpose. In practice, the cloud can just interact with active fog devices.

\subsection{Operational Cost Minimization Problem}

With above-mentioned models, a fog-assisted operational cost minimization problem is formulated as follows,
\begin{subequations}\label{f_12}
\begin{align}
(\textbf{P1})~~&\min~\Gamma_1+\Gamma_2+\Gamma_3+\Gamma_4  \\
s.t.&~\alpha_{i,j}\leq \frac{1}{s_j}(v_{i,j}-s_j/t_j^{\max}),\\
&\sum\nolimits_i\beta_{i,j,k}\leq c_{j,k}\mu_{j,k}-e_{j,k},\\
&(2)-(5),(7)-(8),
\end{align}
\end{subequations}
where $e_{j,k}=\frac{1}{t_j^{\max}-\frac{1}{\mu_{j,k}}}$; (15b) and (15c) are obtained by adjusting the forms of (1) and (6); the decision variables of \textbf{P1} are $\alpha_{i,j}$, $\beta_{i,j,k}$, and $c_{j,k}$.

\section{Algorithm Design}\label{s3}

The objective function and constraints of \textbf{P1} are linear. Moreover, $c_{j,k}$ ($\forall j,~k$) are integer variables. Thus, \textbf{P1} is a mixed integer linear programming (MILP). In addition, the number of constraints and variables is $NJK+2NJ+JK+2J+K$ and $NJK+JK+NJ$, respectively. Since the number of fog devices could be very large, \textbf{P1} is a large-scale mixed integer linear programming (MILP), e.g., when $N=1000$, $J=10$, $K=20$, the number of constraints and variables would be larger than 200000. When solving such a large-scale MILP in a centralized manner, the corresponding computation time would increase dramatically if the problem size becomes large\cite{Chai2016}. Therefore, we are motivated to propose a scalable and distributed algorithm for \textbf{P1} based on PJ-ADMM, which could solve large-scale convex optimization problems efficiently. To solve \textbf{P1} using PJ-ADMM, some transformations are needed. Firstly, we should transform \textbf{P1} into a convex optimization problem by relaxing the constraints related to integer variables. Then, some constraints should be decoupled so that the framework of PJ-ADMM could be used. The transformation detail is given as follows.

\subsection{Problem Transformation}
According to \cite{Neumaier2004}, (15c) could be transformed into $c_{j,k}\geq \frac{1}{\mu_{j,k}}\big(\sum\nolimits_i \beta_{i,j,k}+e_{j,k}\big)$. Since $c_{j,k}\in \mathbb{N}^+$, $c_{j,k}= \lceil\frac{1}{\mu_{j,k}}\big(\sum\limits_i \beta_{i,j,k}+e_{j,k}\big)\rceil$. Thus, $c_{j,k}\leq \frac{1}{\mu_{j,k}}\big(\sum\limits_i\beta_{i,j,k}+e_{j,k}\big)+1$. Taking constraints (6) and (7) into consideration, we have $\sum\nolimits_i \beta_{i,j,k}+e_{j,k}\leq \mu_{j,k}C_{j,k}$. As a result, \textbf{P1} is transformed into \textbf{P2} by discarding some constant items in the objective function,
\begin{subequations}\label{f_14}
\begin{align}
(\textbf{P2})~~&\min~\Gamma  \\
s.t.&~(15\text{b}),(2)-(5)\\
&\sum\limits_i\beta_{i,j,k}\leq \mu_{j,k}C_{j,k}-e_{j,k}, \\
&c_{j,k}= \lceil\frac{1}{\mu_{j,k}}\big(\sum\limits_i \beta_{i,j,k}+e_{j,k}\big)\rceil,
\end{align}
\end{subequations}
where $\Gamma=\sum\limits_i\sum\limits_j \frac{h_iS_iq_{i}\alpha_{i,j}s_jT}{v_{i}}+\sum\limits_i\sum\limits_j\sum\limits_k(\tau_jB_k+\omega_j L_{i,k}T+\frac{\nu_k T}{\mu_{j,k}}(a_{j,k}+b_{j,k}))\beta_{i,j,k}$; the decision variables are $\alpha_{i,j}$ and $\beta_{i,j,k}$. We can first solve \textbf{P2} without considering the last constraint. After obtaining the optimal $\beta_{i,j,k}$, the number of active servers for application $j$ in data center $k$ could be derived based on (16d).

To solve \textbf{P2}, a typical way is to use dual decomposition since the objective function is separable over decision variables. However, the objective function in \textbf{P1} is not strictly convex and dual decomposition could not be used. Otherwise, the Lagrangian would be unbounded below\cite{LiangArxiv2016}. In this paper, we intend to solve \textbf{P3} based on the PJ-ADMM\cite{Deng2014}, which could be used to generate a distributed and parallel algorithm.

When directly applying PJ-ADMM to \textbf{P3}, a centralized algorithm would be incurred since there are couplings among $\beta_{i,j,k}$. Therefore, we continue to transform \textbf{P2}. To avoid the couplings, we adopt a set of auxiliary variables $\beta_{i,j,k}=\gamma_{i,j,k}$, and $\beta_{i,j,k}=l_{i,j,k}$. Consequently, \textbf{P2} could be transformed into \textbf{P3} equivalently,

\begin{subequations}\label{f_15}
\begin{align}
(\textbf{P3})~~&\min~\Gamma  \\
s.t.&~(15\text{b}),(16\text{d}),(2),(4),(5),\\
&\alpha_{i,j}+\sum\nolimits_{k} \gamma_{i,j,k}=\lambda_{i,j},~\forall~i,j,\\
&\gamma_{i,j,k}=\beta_{i,j,k},~\forall i,j,k,\\
&\beta_{i,j,k}=l_{i,j,k},~\forall i,j,k,\\
&\sum\nolimits_il_{i,j,k}\leq \mu_{j,k}C_{j,k}-e_{j,k},~\forall~j,k, \\
&\gamma_{i,j,k}\geq 0,~\forall i,j,k \\
&l_{i,j,k}\geq 0,~\forall i,j,k
\end{align}
\end{subequations}
where the decision variables are $\alpha_{i,k}$, $\gamma_{i,j,k}$, $\beta_{i,j,k}$, and $l_{i,j,k}$.

\subsection{The Proposed Distributed Algorithm}

Define $\boldsymbol{X}$ as the collection of variables $\alpha_{i,k}$, $\gamma_{i,j,k}$, $\beta_{i,j,k}$, and $l_{i,j,k}$ for all $i$,~$j$,~$k$. Denote the augmented Lagrangian of \textbf{P3} by $\mathcal{L}_\rho(\boldsymbol{X};\phi_{i,j},\varphi_{i,j,k},\chi_{i,j,k})$, which is given in \eqref{f_16}, where $\rho$ is the penalty parameter. $\phi_{i,j}$ and $\varphi_{i,j,k}$ are dual variables associated with (17c)-(17e), respectively.

\begin{figure*}[!t]
\normalsize
\setcounter{equation}{17}
\begin{align}\label{f_16}
\mathcal{L}_\rho(\boldsymbol{X};\phi_{i,j},\varphi_{i,j,k},\chi_{i,j,k})= &\Gamma+\sum\limits_i\sum\limits_j \Big(\phi_{i,j}(\alpha_{i,j}+\sum\limits_k \gamma_{i,j,k}-\lambda_{i,j})+\frac{\rho}{2}(\alpha_{i,j}+\sum\limits_k \gamma_{i,j,k}-\lambda_{i,j})^2\Big)\nonumber \\
&+\sum\limits_i\sum\limits_j\sum\limits_k\Big(\varphi_{i,j,k}(\gamma_{i,j,k}-\beta_{i,j,k})+\frac{\rho}{2}(\gamma_{i,j,k}-\beta_{i,j,k})^2\Big) \nonumber \\
&+\sum\limits_i\sum\limits_j\sum\limits_k\Big(\chi_{i,j,k}(\beta_{i,j,k}-l_{i,j,k})+\frac{\rho}{2}(\beta_{i,j,k}-l_{i,j,k})^2\Big)
\end{align}
\hrulefill
\vspace*{4pt}
\end{figure*}

Following the framework of PJ-ADMM, we design a distributed algorithm for \textbf{P3} as follows,

\underline{\textbf{1. Initialization}}: Let all decision variables be zero. For each iteration $w=0,1,2,\cdots$, the following steps are repeated in parallel until convergence.

\underline{\textbf{2.1 $\alpha _{i,j}$-minimization}}: Each fog device $i$ solves the following optimization problem to obtain $\alpha_{i,j}^{w+1}$ in parallel.
\begin{subequations}\label{f_17}
\begin{align}
(\textbf{P4})~\min_{\alpha _{i,j}}~~&\Upsilon_1(\alpha _{i,j},\alpha _{i,j}^w,\gamma_{i,j,k}^w,\phi_{i,j}^w) \\
s.t.~&0\leq \alpha_{i,j}\leq \frac{1}{s_j}(v_{i,j}-s_j/t_j^{\max}),
\end{align}
\end{subequations}
where $\Upsilon_1=\frac{\rho}{2}(\alpha _{i,j}+\sum\limits_k \gamma_{i,j,k}^w-\lambda_{i,j})^2+\frac{\theta_{i,j}}{2}(\alpha _{i,j}-\alpha _{i,j}^w)^2+(\phi_{i,j}^w+\frac{h_iS_iq_{i}s_jT}{v_{i}})\alpha _{i,j}$, where $\theta_{i,j}>0$ ($\forall~i,j$) are the elements of the diagonal matrix used in the proximal term associated with $\alpha_{i,j}$; the solution to \textbf{P4} could be found in Appendix A.

\underline{\textbf{2.2 $\gamma_{i,j,k}$-minimization}}: Each fog device $i$ solves the following optimization problem to obtain $\gamma_{i,j,k}^{w+1}$ in parallel.
\begin{subequations}\label{f_18}
\begin{align}
(\textbf{P5})~\min_{\gamma_{i,j,k}}~~&\Upsilon_2(\gamma_{i,j,k},\gamma_{i,j,k}^w,\alpha _{i,j}^w,\beta _{i,j,k}^w,\phi_{i,j}^w,\varphi_{i,j,k}^w) \\
s.t.~&\gamma_{i,j,k}\geq 0,
\end{align}
\end{subequations}
where $\Upsilon_2=\frac{\rho}{2}(\alpha _{i,j}^w+\sum\limits_k \gamma_{i,j,k}-\lambda_{i,j})^2+\sum\limits_k((\phi_{i,j}^w+\varphi_{i,j,k}^w)\gamma_{i,j,k}+\frac{\rho}{2}(\gamma_{i,j,k}-\beta_{i,j,k}^w)^2+\frac{\sigma_{i,j,k}}{2}(\gamma_{i,j,k}-\gamma_{i,j,k}^w)^2)$, where $\sigma_{i,j,k}>0$ ($\forall~i,j,k$) are the elements of the diagonal matrix used in the proximal term associated with $\gamma_{i,j,k}$; the solution to \textbf{P5} could be found in Appendix B.

\underline{\textbf{2.3 $\beta_{i,j,k}$-minimization}}: Each data center $k$ solves the following optimization problem to obtain $\beta_{i,j,k}^{w+1}$ in parallel.
  \begin{subequations}\label{f_19}
\begin{align}
(\textbf{P6})~\min_{\beta_{i,j,k}}~~&\Upsilon_3(\beta_{i,j,k},\beta_{i,j,k}^w,\gamma_{i,j,k}^w,\varphi_{i,j,k}^w) \nonumber \\
s.t.~&\sum\limits_i\sum\limits_{j} \beta_{i,j,k}s_j\leq A_k^{\max},\\
&\beta_{i,j,k}\geq 0,
\end{align}
\end{subequations}
where $\Upsilon_3=\sum\limits_i\sum\limits_j\Big((\tau_jB_k+\omega_j L_{i,k}T+\frac{\nu_k T}{\mu_{j,k}}(a_{j,k}+b_{j,k})-\varphi_{i,j,k}^w+\chi_{i,j,k}^w)\beta_{i,j,k}+\frac{\rho}{2}(\beta_{i,j,k}-l_{i,j,k}^w)^2+\frac{\rho}{2}(\beta_{i,j,k}-\gamma_{i,j,k}^w)^2+\frac{\eta_{i,j,k}}{2}(\beta_{i,j,k}-\beta_{i,j,k}^w)^2\Big)$, where $\eta_{i,j,k}>0$ ($\forall~i,j,k$) are the elements of the diagonal matrix used in the proximal term associated with $\beta_{i,j,k}$; the solution to \textbf{P6} could be found in Appendix C.

\underline{\textbf{2.4 $l_{i,j,k}$-minimization}}: Each data center $k$ solves the following optimization problem to obtain $l_{i,j,k}^{w+1}$ in parallel.
\begin{subequations}\label{f_20}
\begin{align}
(\textbf{P7})~\min_{l_{i,j,k}}~~&\Upsilon_4(l_{i,j,k},l_{i,j,k}^w,\beta_{i,j,k}^w,\chi_{i,j,k}^w) \nonumber \\
s.t.~&l_{i,j,k}\geq 0,\\
&\sum\limits_il_{i,j,k}\leq \mu_{j,k}C_{j,k}-e_{j,k},
\end{align}
\end{subequations}
where $\Upsilon_4=\sum\limits_i\sum\limits_j\Big(-\chi_{i,j,k}^wl_{i,j,k}+\frac{\rho}{2}(l_{i,j,k}-\beta_{i,j,k}^w)^2+\frac{\kappa_{i,j,k}}{2}(l_{i,j,k}-l_{i,j,k}^w)^2 \Big)$, where $\kappa_{i,j,k}>0$ ($\forall~i,j,k$) are the elements of the diagonal matrix used in the proximal term associated with $l_{i,j,k}$; the solution to \textbf{P7} could be found in Appendix D.

\underline{\textbf{2.5 Dual update}}: Dual variables are updated in the following way, i.e., $\phi_{i,j}^{w+1}=\phi_{i,j}^{w}+\delta \rho (\alpha_{i,j}^{w+1}+\sum\nolimits_k \gamma_{i,j,k}^{w+1}-\lambda_{i,j})$ (where $\delta$ is a positive damping parameter); $\varphi _{i,j,k}^{w+1}=\varphi _{i,j,k}^{w}+\delta \rho(\gamma_{i,j,k}^{w+1}-\beta_{i,j,k}^{w+1})$; $\chi_{i,j,k}^{w+1}=\chi_{i,j,k}^{w}+\delta\rho(\beta_{i,j,k}^{w+1}-l_{i,j,k}^{w+1})$.

\underline{\textbf{3. Termination}}:
If the change of the objective function in two consecutive iterations is lower than the chosen threshold $\varpi$ and feasibility violation metric ($\varpi  = \sum\nolimits_{i = 1}^N {\sum\nolimits_{j = 1}^J {\left| {{\alpha_{i,j}^{w+1}} + \sum\nolimits_{k = 1}^K {{\beta_{i,j,k}^{w+1}}}  - {\lambda _{i,j}}} \right|} }) $ is smaller than $\zeta$, the proposed algorithm would terminate.

 \textbf{\emph{Theorem 1}}
\emph{If the optimal solution set of \textbf{P3} is non-empty, the algorithm developed based on PJ-ADMM would converge to an optimal solution of \textbf{P3} when $\theta_{i,j}>\varsigma$, $\sigma_{i,j,k}>(K+1)\varsigma$, $\eta_{i,j,k}>2\varsigma$, $\kappa_{i,j,k}>\varsigma$, where $\varsigma=\rho(\frac{4}{2-\delta}-1)$, $\rho>0$, $0<\delta<2$.}

\textbf{\emph{Proof:}} In \textbf{P3}, it can be observed that the objective function is separable and continuous over all variables. Moreover, the minimum values of all separable functions are zero and the effective domains of such functions are nonempty/closed. Thus, all separable functions are closed proper convex (i.e., the first assumption for the optimality of PJ-ADMM is satisfied). For the convex optimization problem \textbf{P4}, all inequality constraints are affine. Therefore, the strong duality holds when the optimal solution set is non-empty. Continually, the unaugmented Lagrangian of \textbf{P3} has a saddle point (i.e., the second assumption for the optimality of PJ-ADMM is satisfied). Since the square of spectral norm of the relation matrix associated with each block of variables (there are totally 4 blocks) is 1, $K+1$, 2, and 1, respectively, we can complete the proof according to Lemma 2.2 and Theorem 2.3 in \cite{Deng2014}.

\subsection{Algorithmic Complexity}
In the proposed distributed algorithm, all decisions could be made based on close-form expressions or binary search (see Appendixes A-D). Moreover, all decisions in \textbf{P4}-\textbf{P7} could be made by each entity (e.g., a fog device or a data center) in parallel. Therefore, we analyze the computation time complexity associated with each entity as follows. Let $N_{\text{iter}}$ and $N_b$ be the total number of iterations of the proposed distributed algorithm and the maximum iteration number of binary search (used in solving \textbf{P6} and \textbf{P7}), respectively. Typically, several tens of iterations are needed for binary search\cite{Mou2015}, i.e., $N_b=\mathcal{O}(10)$. The computational time complexity associated each fog device and each data center is given by $\mathcal{O}(N_{\text{iter}}JK)$ and $\mathcal{O}(N_{\text{iter}}NJN_b)$, respectively. Considering the truth that $N\gg K$, the algorithmic computation time mainly depends on the computation time of data centers.

\subsection{Algorithmic Implementation}

\begin{figure}[!htb]
\centering
\includegraphics[scale=0.6]{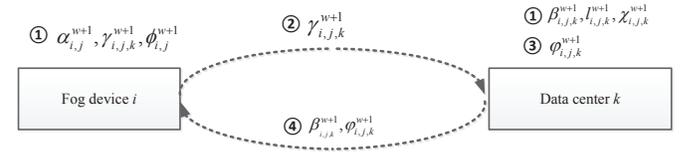}
\caption{An information flow of the proposed distributed algorithm}\label{fig_2}
\end{figure}

The information flow of the proposed distributed algorithm for solving \textbf{P3} could be illustrated by Fig.~\ref{fig_2}. Firstly, each fog device and each data center make their respective decisions in parallel. Then, fog devices broadcast $\gamma_{i,j,k}^{w+1}$ to data centers. After receiving $\gamma_{i,j,k}^{w+1}$, data centers obtain $\varphi_{i,j,k}^{w+1}$ and broadcast $\beta_{i,j,k}^{w+1}$ and $\varphi_{i,j,k}^{w+1}$ to fog devices. In summary, a two-way information broadcast is needed in one iteration. Note that there are communication overheads incurred by transmitting $\gamma_{i,j,k}^{w+1}$, $\beta_{i,j,k}^{w+1}$ and $\varphi_{i,j,k}^{w+1}$ in above-mentioned iteration processes. An alternative way of reducing communication overheads is to implement the proposed distributed algorithm in the cloud, which has powerful computation resources. After obtaining the final results, decision information would be sent back to fog devices. Under this situation, one two-way message interaction is needed.

\section{Performance Evaluation}\label{s4}
\subsection{Simulation Setup}

\begin{table}[!htb]
\renewcommand{\arraystretch}{1.3}
\caption{Parameters related to data centers and applications} \label{tab_1} \centering
\begin{tabular}{|c|c|c|c|c|c|c|}
\hline  $k$     & $j$    & $C_{j,k}$ & $p_{j,k}^\text{idle}$ & $p_{j,k}^\text{peak}$ & $\mu_{j,k}$ & $t_j^{\max}$  \\
\hline
\hline  \multirow{2}{*}{$k=1$}   & $j=1$  & 2000   & 110   & 220  & 3  & 0.5  \\
  & $j=2$  & 1600   & 100    & 200  & 2.625  & 0.6 \\
\hline
\hline  \multirow{2}{*}{$k=2$}   & $j=1$  & 2000   & 95    & 190  & 2.7  & 0.5 \\
& $j=2$  & 1600   & 90    & 180  & 2.4  & 0.6  \\
\hline
\hline  \multirow{2}{*}{$k=3$}   & $j=1$  & 2000   & 120   & 240  & 2.85  & 0.5  \\
& $j=2$  & 1600   & 100   & 200  & 2.25  & 0.6\\
\hline
\end{tabular}
\end{table}

In this section, simulations are conducted to show the performance of the proposed algorithm. For performance comparison, a baseline is adopted, which intends to minimize the operational cost (e.g., energy cost, bandwidth cost and revenue loss) of a cloud service provider without considering the help of fog devices, i.e., $\alpha_{i,j}=0$. The main system parameters are given as follows, $T=1$ hour, $N=1000$, $J=2$, $K=3$, $s_1=0.25$ Mb, $s_2=0.5$ Mb, $A_1^{\max}=10^5$ Mbps, $A_2^{\max}=0.9\times 10^5$ Mbps, $A_3^{\max}=0.8\times 10^5$ Mbps, $B_1=0.005$ \$/Mbps/hour\cite{Greenberg2009}, $\text{PUE}_1=1.13$, $\text{PUE}_2=1.14$, $\text{PUE}_3=1.15$\cite{Cavdar2012}, $\tau_j=1$ Mb, $\nu_1=30$, $\nu_2=35$, $\nu_3=40$ (in \$/MWh). Other parameters related to data centers and applications are provided in Table~\ref{tab_1}\cite{Lei2012}. Suppose that $v_{i,j}$ follows a uniform distribution with parameters 2.25 and 3 (in requests/second), i.e., $v_{i,j}\sim \mathcal{U}(2.25,3)$. Since fog devices would consume higher power compared with cloud servers given the same service capability as in \cite{DengR2016}, we set $q_{i}^{\text{peak}}\sim \mathcal{U} (440,500)$ (in Watts). In addition, we assume that $S_i\sim \mathcal{U}(30,60)$, $\lambda_{i,j}\sim \mathcal{U}(\frac{1}{2N}\sum\nolimits_kC_{j,k}\mu_{j,k}, \frac{1}{N}\sum\nolimits_kC_{j,k}\mu_{j,k})$ (in requests/second), $L_{i,k}\sim \mathcal{U}(10,40)$ (in ms). We set $q_{i}^{\text{idle}}=0.5 q_{i}^{\text{peak}}$\cite{Kontorinis2012}, $\delta=1$.

\subsection{Simulation Results}

\begin{figure*}
\centering
\subfigure[Total cost]{
\begin{minipage}[b]{0.315\textwidth}
\includegraphics[width=1\textwidth]{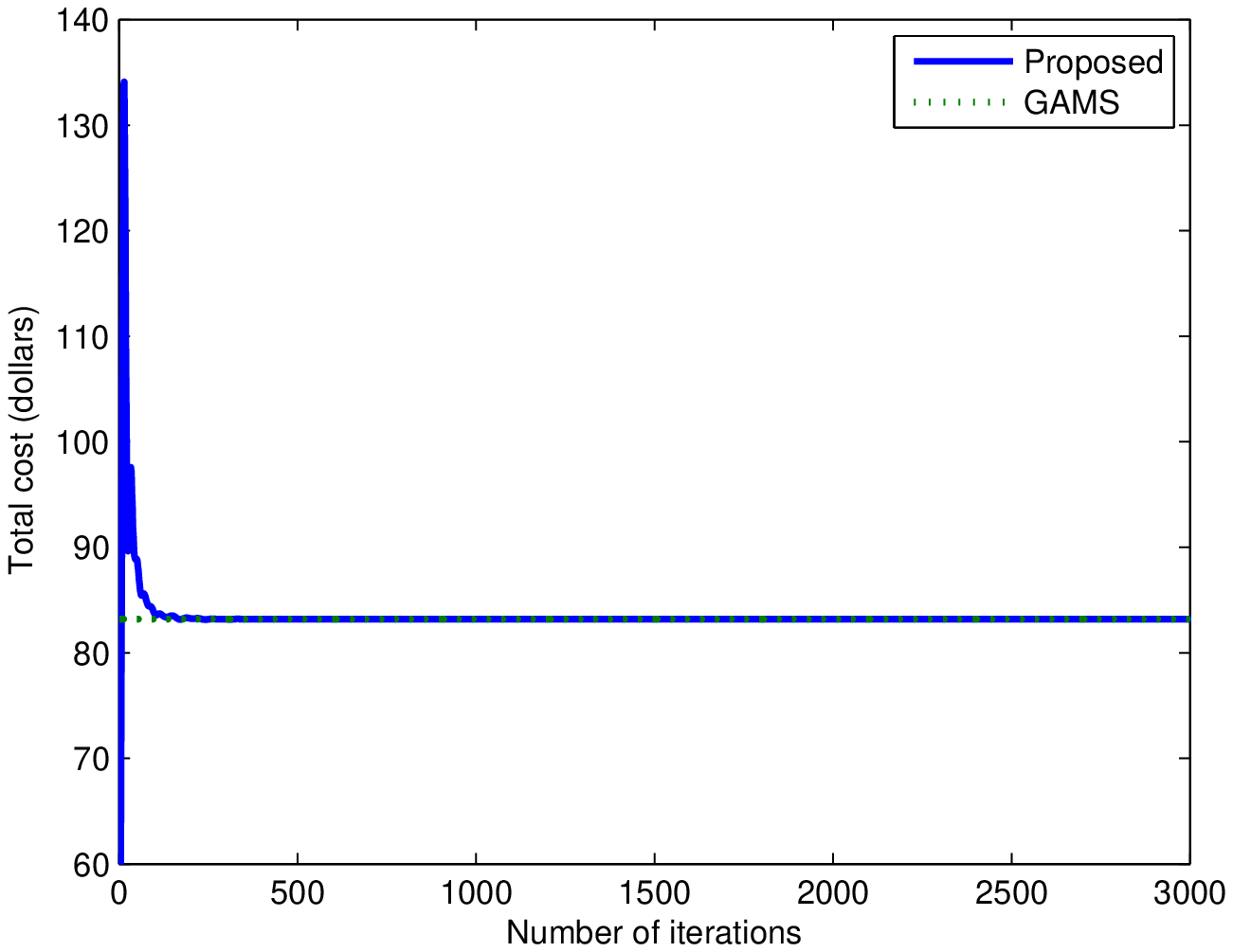}
\end{minipage}
}
\subfigure[Primal residual]{
\begin{minipage}[b]{0.315\textwidth}
\includegraphics[width=1\textwidth]{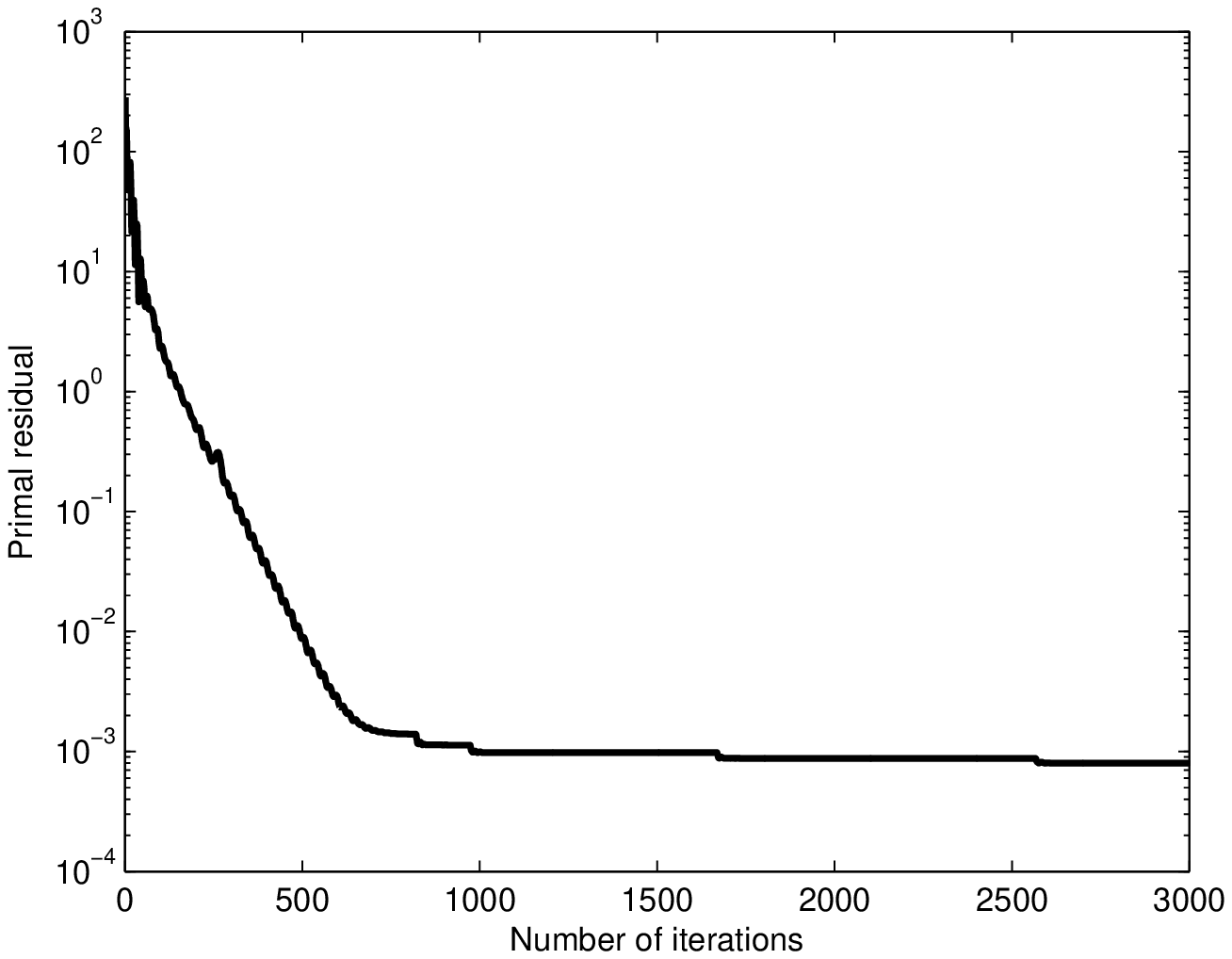}
\end{minipage}
}
\subfigure[Feasibility metric]{
\begin{minipage}[b]{0.315\textwidth}
\includegraphics[width=1\textwidth]{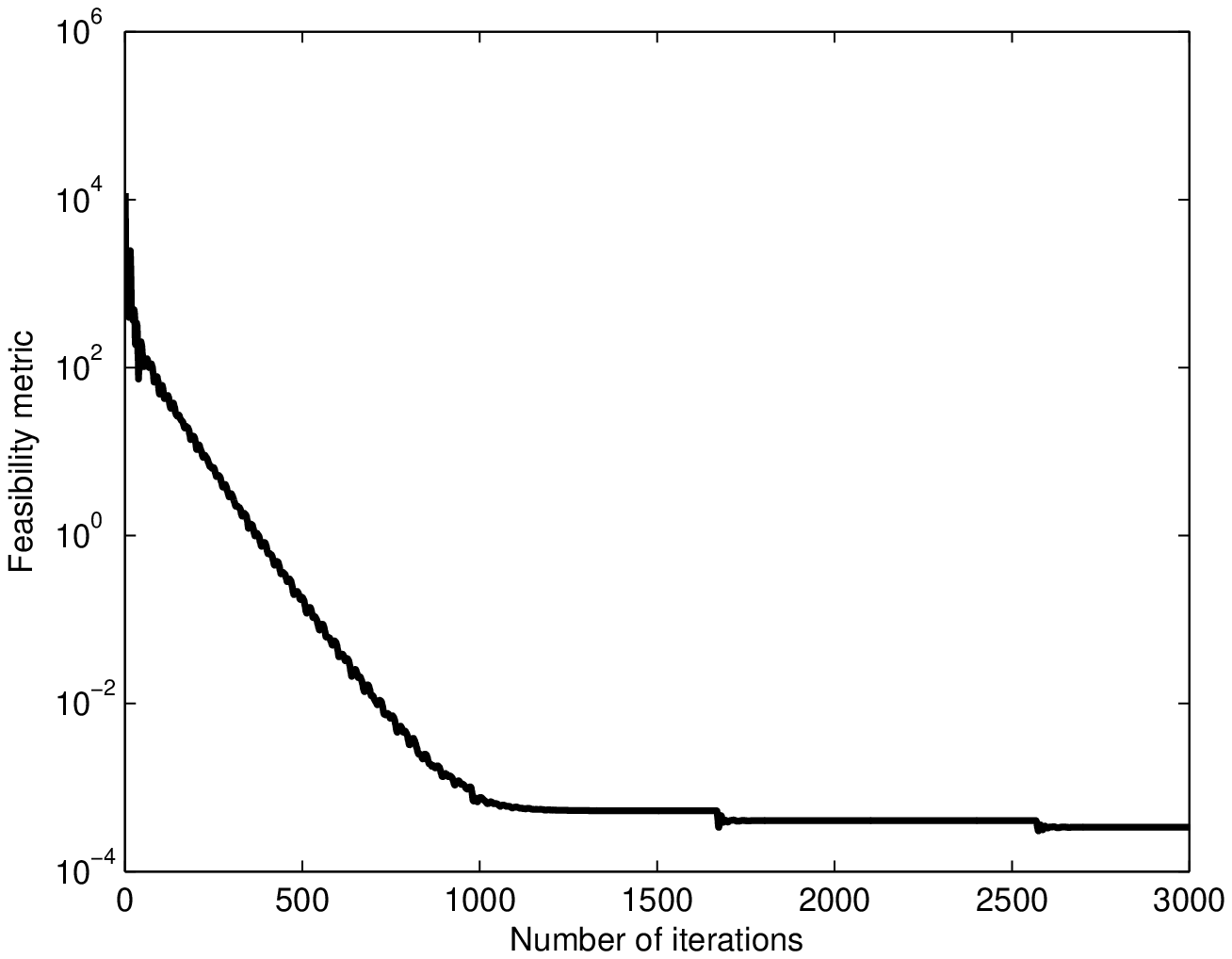}
\end{minipage}
}
\caption{Convergence results of the proposed algorithm ($\rho=0.002$)} \label{fig_3}
\end{figure*}

\begin{figure*}
\centering
\subfigure[Total cost]{
\begin{minipage}[b]{0.315\textwidth}
\includegraphics[width=1\textwidth]{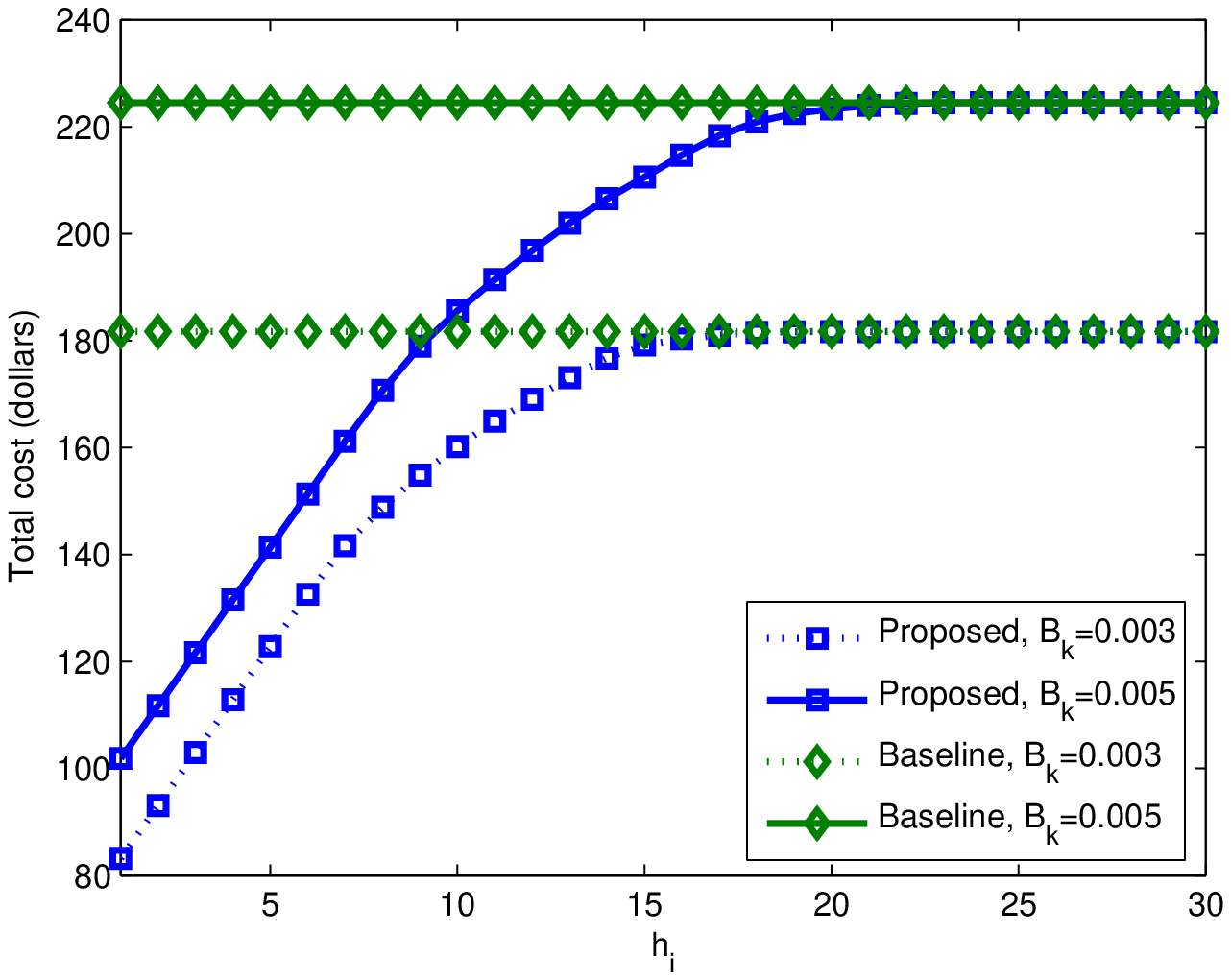}
\end{minipage}
}
\subfigure[Workload finished by fog devices]{
\begin{minipage}[b]{0.315\textwidth}
\includegraphics[width=1\textwidth]{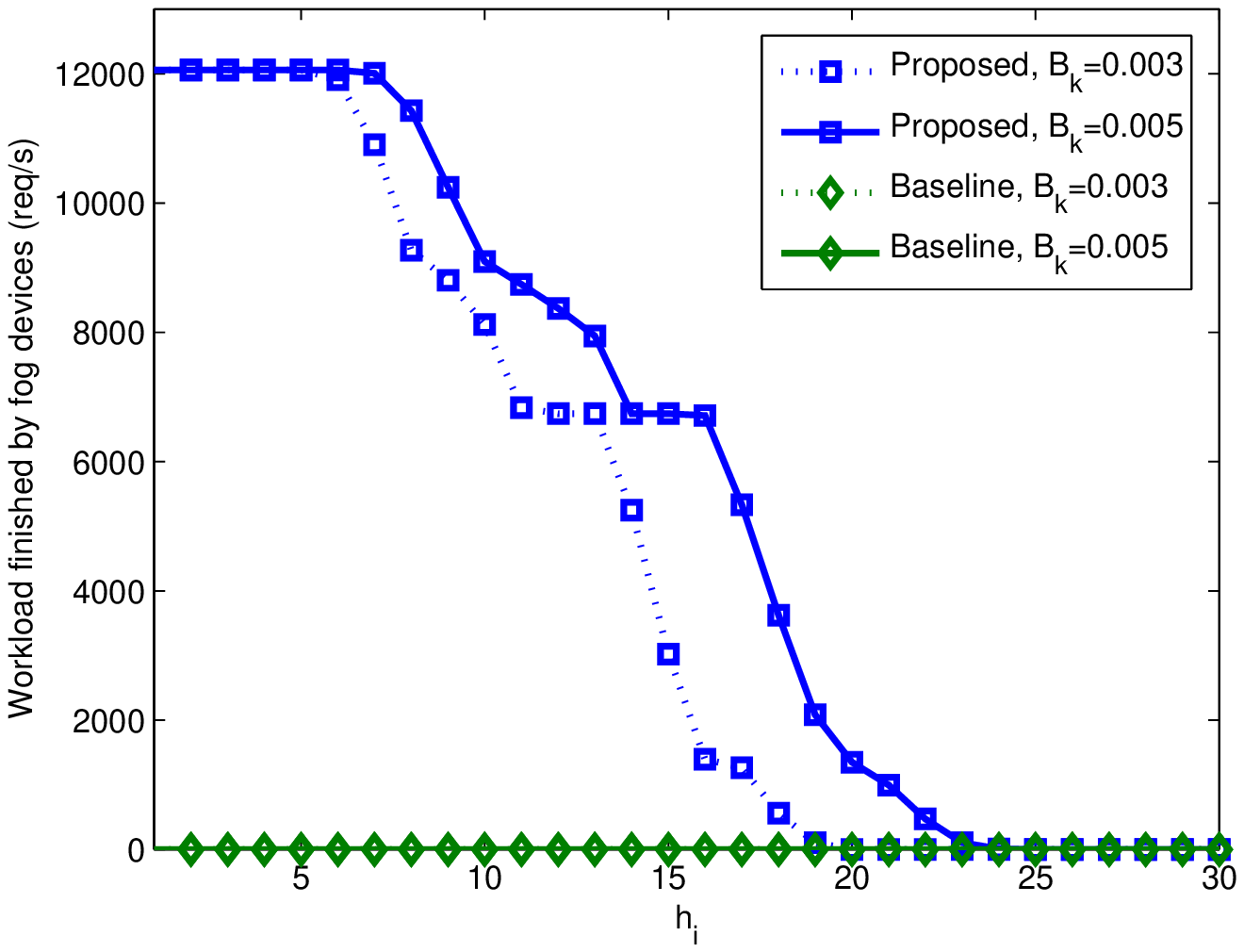}
\end{minipage}
}
\subfigure[Workload finished by the Cloud]{
\begin{minipage}[b]{0.315\textwidth}
\includegraphics[width=1\textwidth]{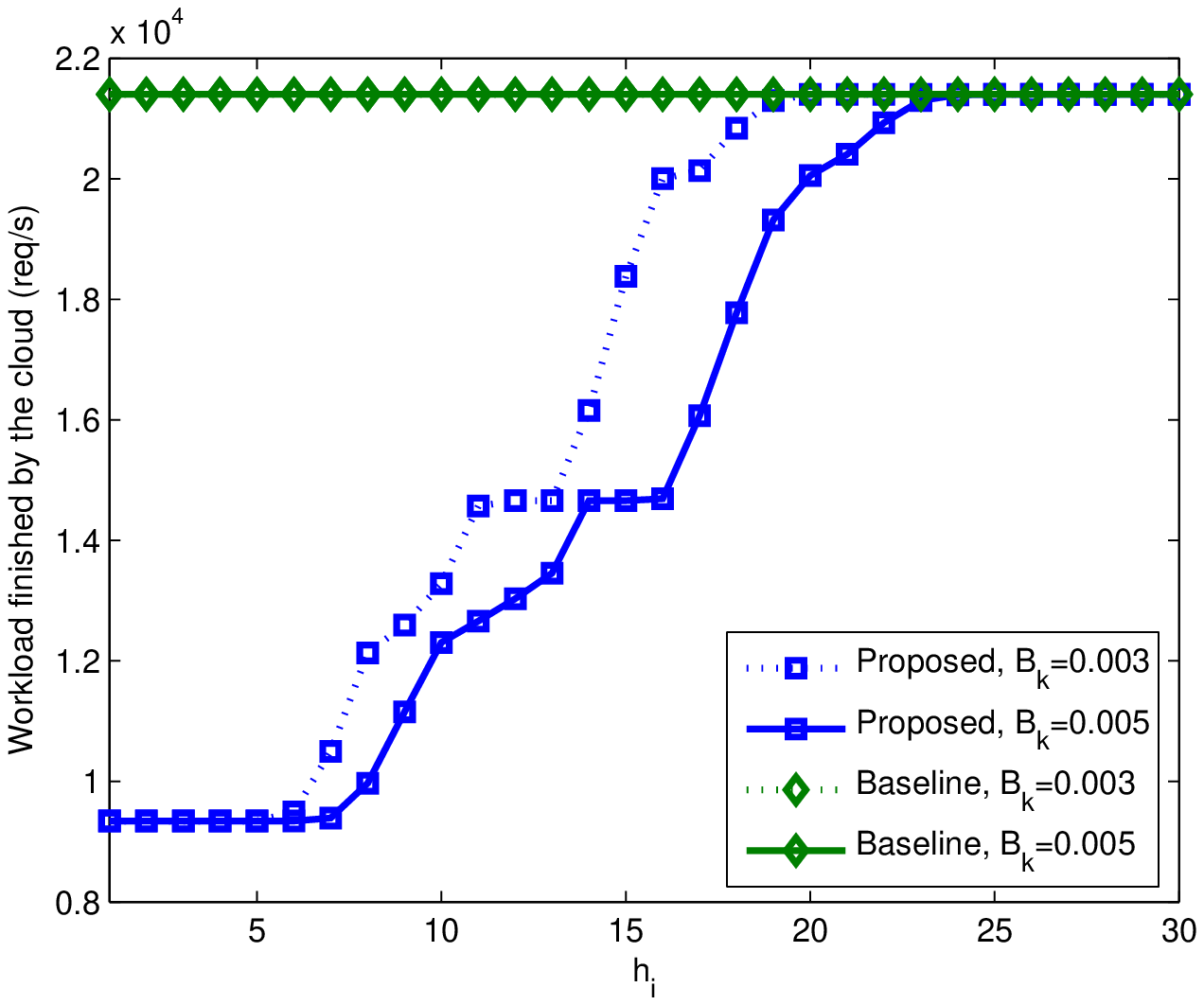}
\end{minipage}
}
\caption{Performances of the proposed algorithm under varying $h_i$ and $B_k$ ($\omega_j=3\times10^{-8}$)} \label{fig_4}
\end{figure*}

\begin{figure*}
\subfigure[Total cost]{
\begin{minipage}[b]{0.315\textwidth}
\includegraphics[width=1\textwidth]{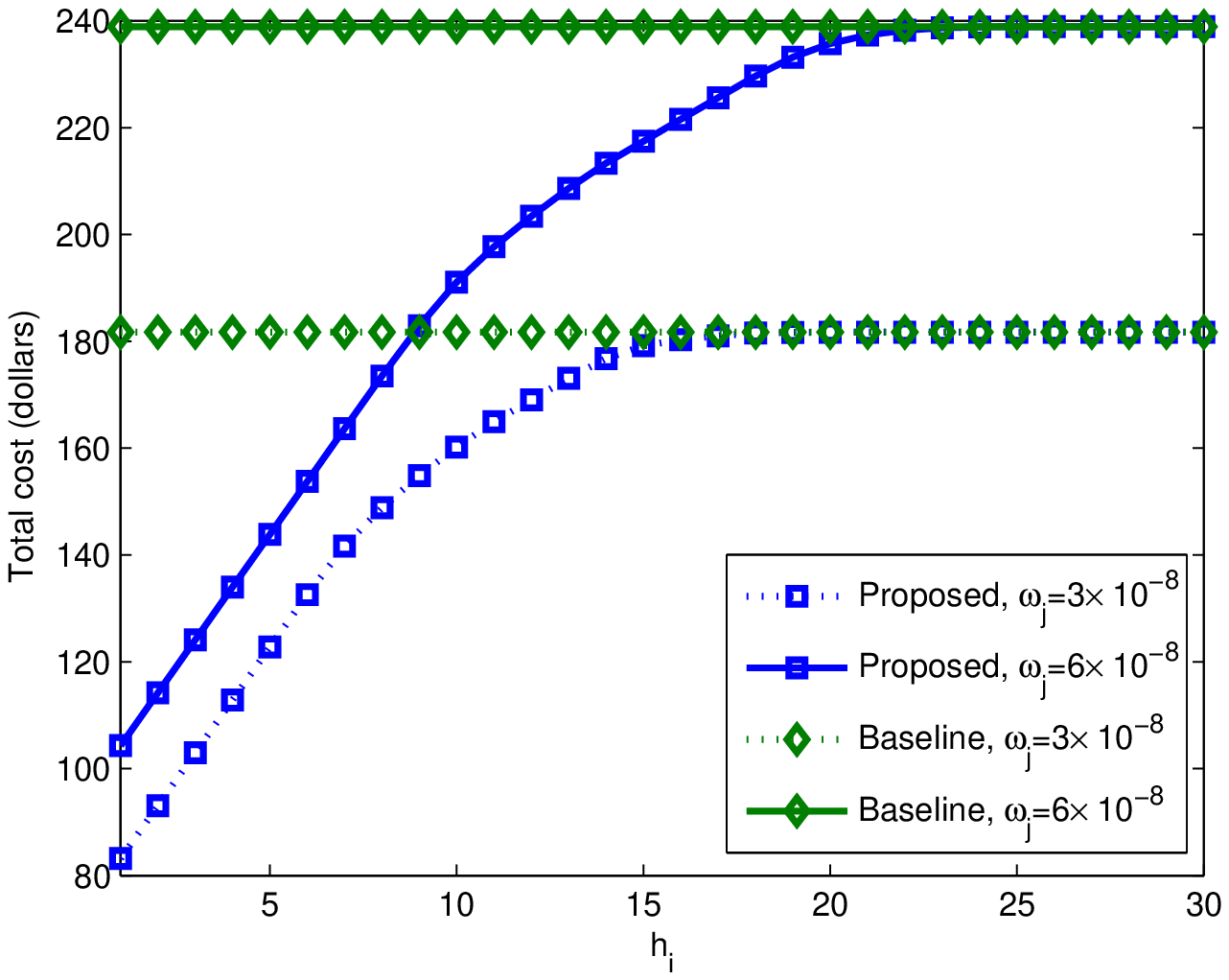}
\end{minipage}
}
\subfigure[Workload finished by fog devices]{
\begin{minipage}[b]{0.315\textwidth}
\includegraphics[width=1\textwidth]{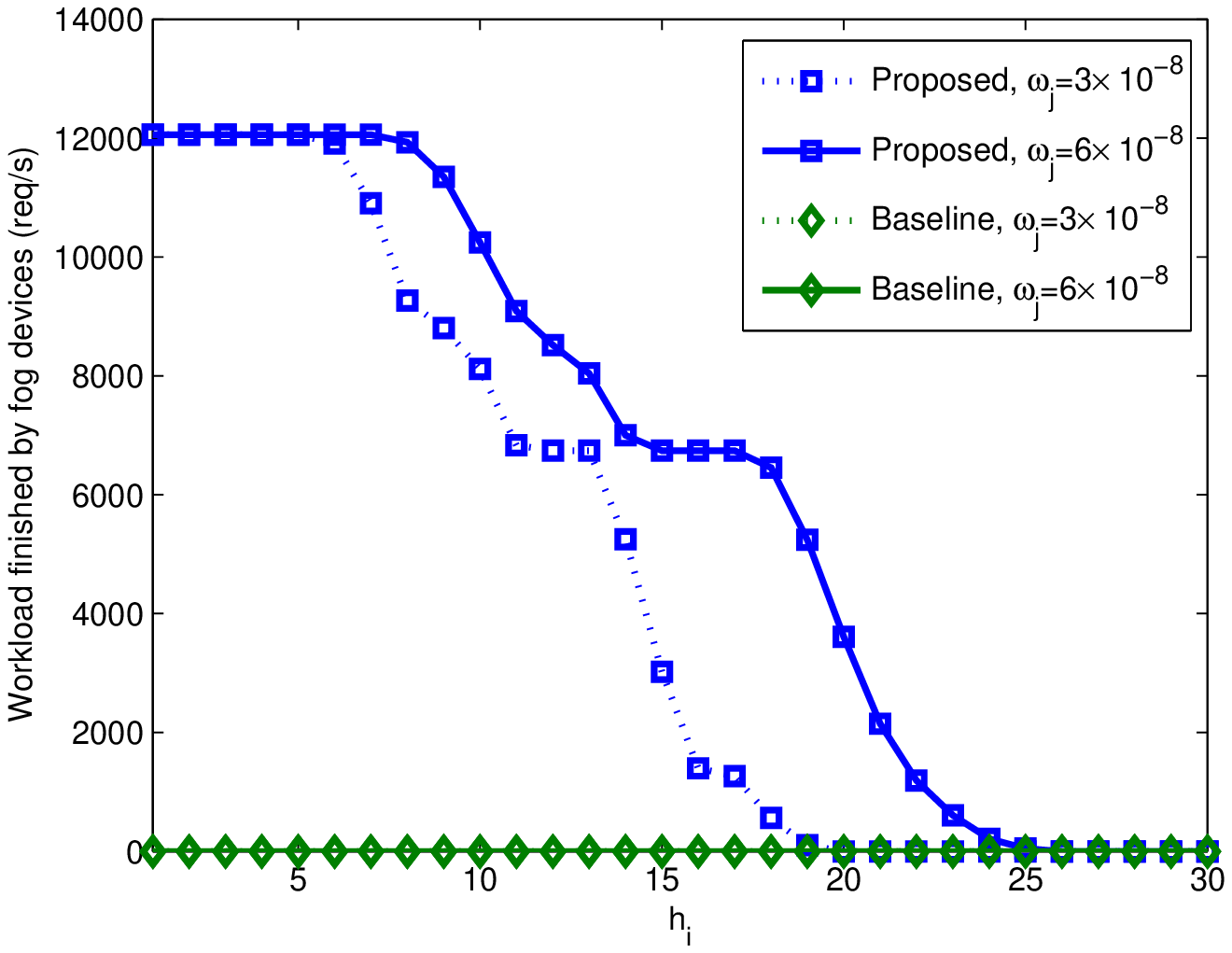}
\end{minipage}
}
\subfigure[Workload finished by the Cloud]{
\begin{minipage}[b]{0.315\textwidth}
\includegraphics[width=1\textwidth]{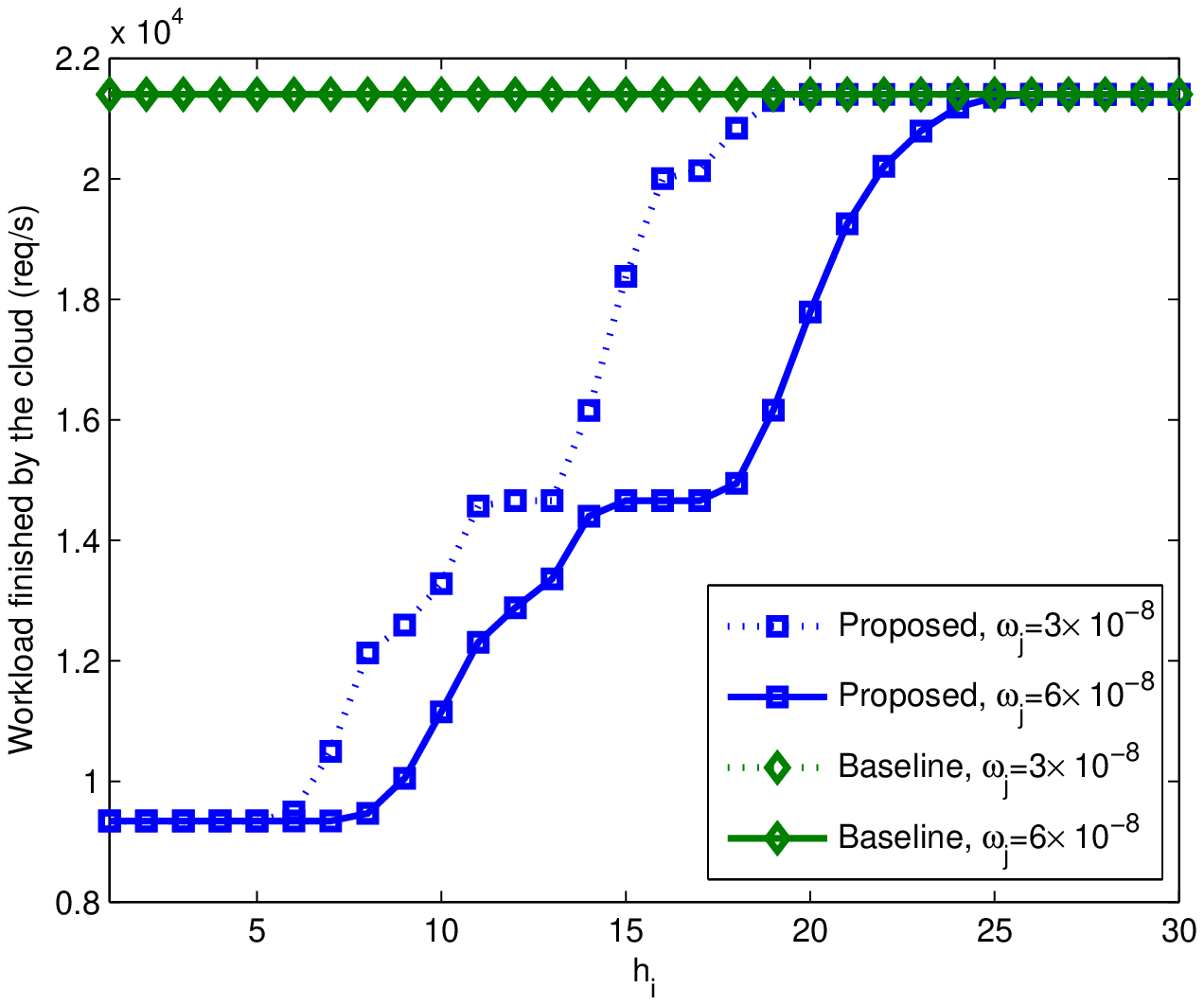}
\end{minipage}
}
\caption{Performances of the proposed algorithm under varying $h_i$ and $\omega_j$ ($\beta=0.003$)} \label{fig_5}
\end{figure*}

\subsubsection{Convergence results of the proposed algorithm}
Since \textbf{P2} is a relaxed problem of \textbf{P1}, there may be a performance gap between the optimal total cost generated by the proposed algorithm and the optimal solution of \textbf{P1}. To obtain the optimal solution of \textbf{P1}, GAMS commercial solver\footnote{http://www.gams.com/} is adopted. In Fig.~\ref{fig_3}, it can be observed that the above-mentioned performance gap is negligible, e.g., the relative optimality gap is 0.0094\% in this scenario. Although several thousands of iterations are needed to find the optimal solution of \textbf{P2}, it does not mean that the proposed PJ-ADMM-based algorithm is inefficient. In fact, the proposed algorithm could find a suboptimal solution with very low optimality loss in several hundreds of iterations. For example, running on a single Intel Core i5-2410M 2.3GHz server with 4G RAM, the relative optimality loss achieved by the proposed algorithm at iteration 600 is smaller than 0.00006\% and the time consumed is 42 seconds, which would become 14 seconds if a parallel implementation is considered (note that there are at least $K$=3 parallel optimization subproblems in each step of the distributed algorithm).

\subsubsection{The impacts of $h_i$, $B_k$ and $\omega_j$}
Since $h_i$ represents the economical compensation factor for fog devices, changing the value of $h_i$ would affect the extent of their participation. Specifically, fog devices would process less workloads with the increase of $h_i$ as shown in Fig.~\ref{fig_4}. According to workload balance, the total workload finished by the cloud would increase with the increase of $h_i$. Under the current simulation scenario, a win-win situation for fog devices and the cloud can be achieved given $1<h_i<20$ when $B_k=0.005$. In addition, given the same $h_i$, the number of requests processed by the cloud is increasing with the decrease of $B_k$, which is obvious since $B_k$ could be regarded as the ``weight" of $\beta_{i,j,k}$ in the objective function of \textbf{P1}. In the future, $B_k$ will continue to decrease, which does not mean that fog computing is unattractive in IoT era. The reason is that some time-critical IoT applications have to be processed locally, e.g., health monitoring. Therefore, more and more requests would be processed by fog devices with the increase of $\omega_j$ as shown in Fig.~\ref{fig_5}(b).

\subsubsection{Cost components under different algorithms}
\begin{figure}[!htb]
\centering
\includegraphics[scale=0.52]{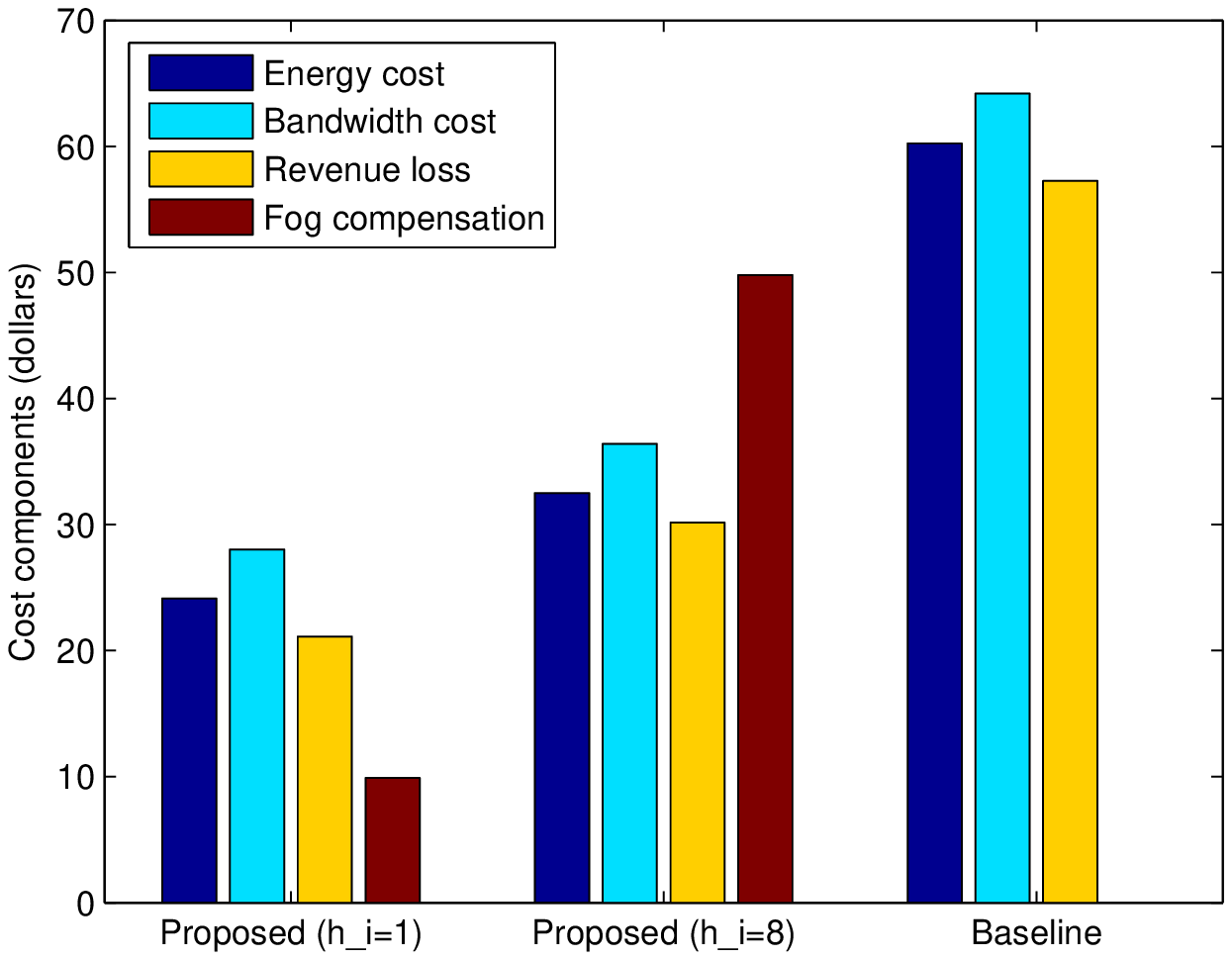}
\caption{Cost components ($B_k=0.003$ and $\omega_j=3\times 10^{-8}$)}\label{fig_6}
\end{figure}

In Fig.~\ref{fig_6}, we provide cost components under the proposed algorithm and the baseline. It can be observed that fog devices can help the cloud to reduce the operational cost by processing nearby user requests, resulting in lower energy cost, bandwidth cost and revenue loss, e.g., relative cost reduction (RCR) achieved by the proposed algorithm is 54.22\% and 18.08\% when $h_i=1$ and $h_i=8$, respectively. Though different parameter configurations would affect the value of RCR as shown in next subsection, the performance of the proposed algorithm is not worse than the baseline. The reason is that the proposed algorithm would be equivalent to the baseline if $h_i$ required by each fog device is too high.

\begin{figure}[!htb]
\centering
\includegraphics[scale=0.52]{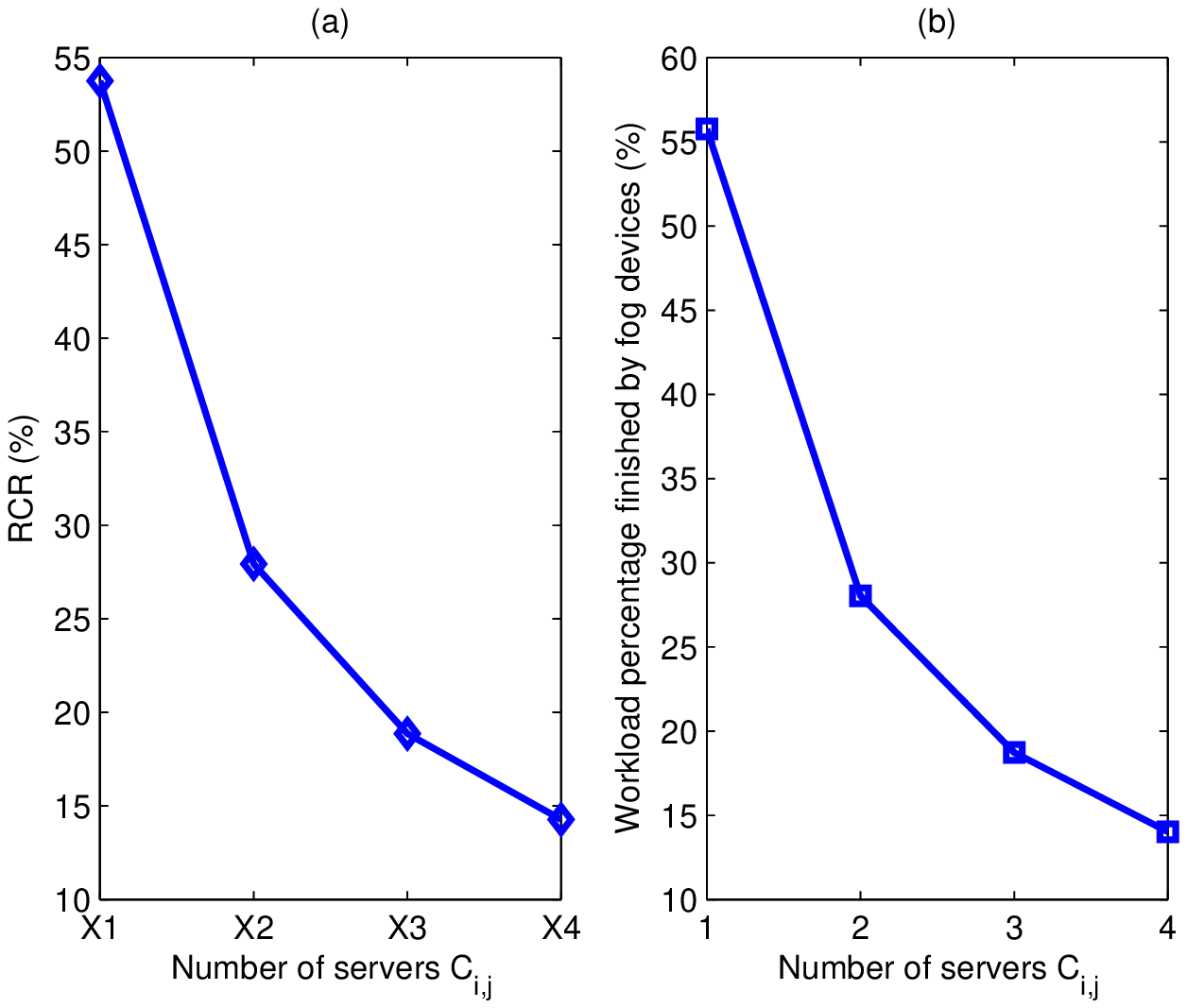}
\caption{Relative cost reduction and workload percentage of fog devices}\label{fig_7}
\end{figure}

\subsubsection{The impact of workloads}
We modify the workload $\lambda_{i,j}$ by enlarging $C_{i,j}$ several times (e.g., 1,2,3,4), considering that $\lambda_{i,j}\sim \mathcal{U}(\frac{1}{2N}\sum\nolimits_kC_{j,k}\mu_{j,k}, \frac{1}{N}\sum\nolimits_kC_{j,k}\mu_{j,k})$. Since the total processing capacity of fog devices is limited, more workloads would be dispatched to the cloud for processing with the increase of incoming workloads at fog devices. As a result, the ratio of fog workload to the total workload and RCR would be reduced simultaneously as shown in Fig.~\ref{fig_7}. In the future, more and more fog devices will be deployed to support some emerging time-critical IoT applications, resulting in larger processing capacity of fog devices and increasing RCR.

\section{Conclusions}\label{s5}
In this paper, we investigated the problem of reducing the operational cost of geo-distributed cloud data centers with the help of fog devices. To be specific, we first formulated a fog-assisted operational cost minimization problem for a cloud service provider with the consideration of economical compensation paid to fog devices. Then, we proposed a parallel and distributed load balancing algorithm to solve the formulated problem based on PJ-ADMM. Note that the proposed algorithm has low computational complexity since all decisions could be obtained based on close-form expressions or binary search. Extensive simulation results showed the effectiveness of the proposed algorithm.

\appendices

\section{The solution to \textbf{P4}}
\begin{IEEEproof}
Let the gradient of $\Upsilon_1$ with respect to $\alpha_{i,j}$ be zero, we have $\nabla _{\alpha_{i,j} } \Upsilon_1$=0. Let $\theta_{i,j}=\bar{\theta}$ for all $i,~j$, we have $\alpha_{i,j}^*=\frac{\rho(\lambda_{i,j}-\sum\limits_k r_{i,j,k}^w)+\bar{\theta}\alpha_{i,j}^w-\phi_{i,j}^w-\frac{h_iS_iq_{i}s_jT}{v_{i}}}{\rho+\bar{\theta}}$. Note that $\alpha_{i,j}\in [0,~(v_{i,j}/s_j-1/t_j^{\max})]$, we have $\alpha_{i,j}^{w+1}=\min(\max\{0,\alpha_{i,j}^*\},(v_{i,j}/s_j-1/t_j^{\max}))$.
\end{IEEEproof}

\section{The solution to \textbf{P5}}
\begin{IEEEproof}
Let the gradient of $\Upsilon_2$ with respect to $\gamma_{i,j,k}$ be zero,, we have $\nabla _{\gamma_{i,j,k} } \Upsilon_2$=0. Let $\sigma_{i,j,k}=\bar{\sigma}$ for all $i,~j,~k$, we have $\gamma_{i,j,k}^*=\frac{\frac{\rho\sum\nolimits_{k}y_{i,j,k}}{(K+1)\rho+\bar{\sigma}}-y_{i,j,k}}{(\rho+\bar{\sigma})}$, where $y_{i,j,k}=\rho(\alpha_{i,j}^w-\beta_{i,j,k}^w-\lambda_{i,j})+(\phi_{i,j}^w+\varphi_{i,j,k}^w-\gamma_{i,j,k}^w\bar{\sigma})$. Note that $\gamma_{i,j,k}\geq 0$, we have $\gamma_{i,j,k}^{w+1}=\max\{0,\gamma_{i,j,k}^*\}$.
\end{IEEEproof}

\section{The solution to \textbf{P6}}
\begin{IEEEproof}
Let the first derivative of $\Upsilon_3$ with respect to $\beta_{i,j,k}$ be zero, we have
 \begin{align} \label{f_a1}
\beta_{i,j,k}^*=\frac{g_{i,j,k}}{(\eta_{i,j,k}+2\rho)},
\end{align}
where $g_{i,j,k}=\rho(\gamma_{i,j,k}^w+l_{i,j,k}^w)+\eta_{i,j,k}\beta_{i,j,k}^w-(\tau_jB_k+\omega_jL_{i,k}T+\frac{\nu_kT}{\mu_{j,k}}(a_{j,k}+b_{j,k})+\chi_{i,j,k}^w-\varphi_{i,j,k}^w)$. Denote the set $\{(i,j)|g_{i,j,k}<0\}$ by $\mathcal{I}_k$. If $(i,j)\in \mathcal{I}_k$, then, $\beta_{i,j,k}^{w+1}=0$. If $\sum\nolimits_{(i,j)\notin \mathcal{I}_k}\beta_{i,j,k}^*s_j \leq A_k^{\max}$, then, $\beta_{i,j,k}^{w+1}=\beta_{i,j,k}^*$ for all $(i,j)\notin \mathcal{I}_k$. Otherwise, for all $(i,j)\notin \mathcal{I}_k$, using KKT optimality conditions, we have
 \begin{align} \label{f_a2}
\beta_{i,j,k}^*=\frac{g_{i,j,k}-\varrho_k\tau_j}{(\eta_{i,j,k}+2\rho)},
\end{align}
where $\varrho_k (\forall k)$ are non-negative dual variables related to (21a); the optimal $\varrho_k (\forall k)$ are decided by the equation $\sum\nolimits_{(i,j)\notin \mathcal{I}_k}\beta_{i,j,k}\tau_j=A_k^{\max}$. Since $\beta_{i,j,k}^*$ is gradually reduced to be zero with the increase of $\varrho_k$ according to \eqref{f_a2}, we can find the optimal $\varrho_k$ based on binary search\cite{Liang2016TSG}.
\end{IEEEproof}

\section{The solution to \textbf{P7}}
\begin{IEEEproof}
Let $\kappa_{i,j,k}=\bar{\kappa}$ for all $i,~j,~k$. Since the first derivative of $\Upsilon_4$ with respect to $l_{i,j,k}$ should be zero, we have
\begin{align} \label{f_a3}
l_{i,j,k}^*=\frac{z_{i,j,k}}{(\rho+\bar{\kappa} )},
\end{align}
where $z_{i,j,k}=\bar{\kappa}l_{i,j,k}^w+\rho \beta_{i,j,k}^w+\chi_{i,j,k}^w$. Denote the set $\{i|z_{i,j,k}<0\}$ by $\mathcal{K}_{j,k}$ ($\forall~j,~k$). If $i\in \mathcal{K}_{j,k}$, then, $l_{i,j,k}^{w+1}=0$. If $\sum\nolimits_{i\notin \mathcal{K}_{j,k}}l_{i,j,k}\leq \mu_{j,k}C_{j,k}-e_{j,k}$, then, $l_{i,j,k}^{w+1}=l_{i,j,k}^*$ for all $i\notin \mathcal{K}_{j,k}$. Otherwise, for all $i \notin \mathcal{K}_{k}$, using KKT optimality conditions, we have
 \begin{align} \label{f_a4}
l_{i,j,k}^*=\frac{z_{i,j,k}-\xi_{j,k}}{(\rho+\bar{\kappa})},
\end{align}
where $\xi_{j,k} (\forall~j,~k)$ are non-negative dual variables related to (22b); the optimal $\xi_{j,k}$ is decided by the equation $\sum\nolimits_{i\notin \mathcal{K}_{j,k}}l_{i,j,k}=\mu_{j,k}C_{j,k}-e_{j,k}$. Since $l_{i,j,k}^*$ is gradually reduced with the increase of $\xi_{j,k}$ according to \eqref{f_a4}, we can find the optimal $\xi_{j,k}$ based on binary search.
\end{IEEEproof}

\end{document}